\documentclass[useAMS,usenatbib,usegraphicx]{mn2e}

\def\lsim{\la}

\def\text{\rm}
\def\Brunt{Brunt-V\"{a}is\"{a}l\"{a}\ }
\def\arccot{{}\mathrm{arccot}{}}
\def\vector#1{{\bmath{#1}}}

\title[Radiative zone solar magnetic fields and g-modes]%
{Radiative zone solar magnetic fields and g-modes}

\author[T. I. Rashba, V. B. Semikoz and J. W. F. Valle]%
{T. I. Rashba$^{1,2}$\thanks{E-mail:timur@mppmu.mpg.de (TIR)}, 
V. B. Semikoz$^{2,3}$\thanks{E-mail:semikoz@ific.uv.es (VBS)}
and 
J. W. F. Valle$^{3}$\thanks{E-mail:valle@ific.uv.es (JWFV)}\\
$^{1}$Max-Planck-Institut f\"ur Physik (Werner-Heisenberg-Institut),
F\"ohringer Ring 6, D-80805 M\"unchen, Germany\\
$^{2}$Pushkov Institute of Terrestrial Magnetism, Ionosphere and
Radiowave Propagation of the Russian Academy of Sciences,\\ 
IZMIRAN, Troitsk, Moscow region, 142190, Russia\\
$^{3}$AHEP Group, Instituto de F\'isica Corpuscular -- C.S.I.C./Universitat
de Val\`encia,\\ 
Edificio Institutos de Paterna, Apt 22085, E-46071, Val\`encia, Spain}

\begin{document}

\date{\today}

\pagerange{\pageref{firstpage}--\pageref{lastpage}} \pubyear{2006}

\maketitle

\label{firstpage}

\begin{abstract}
  We consider a generalized model of seismic-wave propagation that
  takes into account the effect of a central magnetic field in the
  Sun. We determine the g-mode spectrum in the perturbative magnetic
  field limit using a one-dimensional Magneto-Hydrodynamics (MHD)
  picture. We show that central magnetic fields of about 600-800~kG
  can displace the pure g-mode frequencies by about 1\%, as hinted by
  the helioseismic interpretation of GOLF observations.
\end{abstract}


%\preprint{IFIC/05-62; MPP-2005-151}

\section{Introduction}

Currently there is very little direct information about the structure
and strength of magnetic fields in the radiative zone (RZ) of the Sun,
for a short review see Introduction of the paper \citep{Burgess04a}.
Some authors argue that for the young Sun ($\lsim 3-30$~Myr)
relatively small fields, $\lsim 10$~kG~\citep{Moss03} and $\lsim
1$~Gauss~\citep{Kitchatinov01} could survive, being relic fields
captured from the primordial ones in the protostar plasma.  For the
Sun at the present epoch there is an upper bound of $2-3$~MG near the
tachocline obtained from the magnetic splitting of acoustic
oscillations~\citep{Ruzmaikin02}.  However, some authors have
considered very strong magnetic fields in the RZ, up to 30
MG~\citep{Couvidat03}.

Here we suggest a new way to estimate the magnetic field strengths in
the RZ of the Sun by relating them to the frequency shifts of g-mode
candidates suggested by the first observations made with the GOLF
(Global Oscillations at Low Frequencies) experiment~\citep{TC04}.  We
discuss some effects of RZ magnetic fields which could explain the
displacement of g-mode frequencies with respect to the theoretical
frequencies calculated in the absence of magnetic field. Indeed the
existence of such shifts are hinted in GOLF's data.  If eventually
confirmed by further data, the idea that RZ magnetic fields cause such
frequency shifts would provide us with a useful tool to estimate their
magnitude.

In order to find spectra of seismic waves accounting for the magnetic
field in the RZ a number of assumptions is required. For example:
\begin{enumerate}
\item
We consider ideal MHD neglecting both the heat conductivity and viscosity
contributions to energy losses, as well as the ohmic dissipation.
\item
We linearize the MHD equations about a static background
configuration, {\it i.e.} a background configuration which is time
independent and for which the background fluid velocity vanishes,
$\vector{v}_0 = 0$.
\item
We assume the fluctuations to be adiabatic, with the contributions of
fluctuations to the heat source vanishing: $Q'=0$.
\item Moreover, we consider a fully ionized ideal gas, so that the
  thermodynamic quantity, first adiabatic exponent $\gamma = c_p/c_V$,
  is time independent and uniform. For numerical estimates we will
  take $\gamma=5/3$ for hydrogen plasma.
\item
We adopt the Cowling approximation, which amounts to the neglect
of perturbations of the gravitational potential, ({i.e.:} $\phi' =
0$).
\item
We assume a rectangular geometry with Cartesian coordinates: $x$,
$y$, and $z$, where $z$ corresponds to the solar radial direction.
The background quantities vary along the $z$ direction only (which
implies the local gravitational acceleration, ${\bf g}$, is directed
along the $z$ axis, but in opposite direction). We also take a
constant, uniform background magnetic field, ${\bf B}_0$, pointing
along the $x$ axis.
\item
The background mass-density profile is assumed to be exponential,
$\rho_0 = \rho_c \, \exp[- z/H]$, for constant $\rho_c$ and $H$.
The conditions of hydrostatic equilibrium for the background then
determine the profiles of thermodynamic quantities, and in
particular imply $\gamma$ is a constant.
We assume that the \Brunt frequency is zero in the convective zone (CZ)
and non-zero, but constant in the RZ.
\end{enumerate}

In what follows, we shall again specify the assumptions used, as they
are needed, in order to keep clear which results rely on which
assumptions.

Note that, deep within the radiative zone, the last approximation
above holds to very good approximation for real mass-density profiles
obtained by Standard Solar Models, provided we identify the $z$
direction with the radial direction.
The constancy of $\gamma$ in this region is also expected since the
highly-ionized plasma satisfies an ideal gas equation-of-state to good
approximation.  The rectangular geometry provides a reasonable
approximation so long as we do not examine too close to the solar
centre. What is important about our choice for ${\bf B}_0$ is that it
is slowly varying in the region of interest, and it is perpendicular
to both ${\bf g}$ and all background gradients, $\nabla \rho_0$,
$\nabla p_0$, {\it etc}.

As suggested in~\citep{Burgess04a} such one-Dimensional (1-D) picture
can be fully described in analytical terms in contrast to the 3-D
case. There are two parameters which describe the spectra of
magneto-gravity waves~\citep{Burgess04a}: (i) strength of the
background magnetic field $B_0$ and (ii) the dimensionless transversal
wave number $K=k_x H$. Here $H$ is the density scale height and $k_x$
is the projection of the wavevector onto the x axis. Let us
estimate the value of the transversal wave number that could be
relevant for the g-mode candidates observed on the photosphere.

Since g-modes decay in the CZ as $\sim e^{-Kz/H}$,
only modes with low transversal wave number $K\sim 1-4$ (long wave
lengths) could be seen at the photosphere.  This follows from the
simple estimate for the longitudinal fluid velocity $v_z(z)$ which is
directed along the Sun-Earth line and causes the Doppler shifts of
optic lines registered by the GOLF experiment:
\begin{equation}
\label{eq:vz} 
v_z(z=R_{\odot})=\frac{NH}{K}\frac{b_z}{B_0}e^{-3K}\simeq
2~\rm{mm~sec^{-1}}.
\end{equation} 
This formula comes from Eq.~(14) (equivalent to our Eq.~(\ref{B_z}))
and Eq.~(30) of Ref.~\citep{Burgess04a} for the decaying solution
$B_z^{(1)}(z)=b_z e^{-Kz/H}$, where $B_z^{(1)}$ is the $z$-component
of the magnetic field perturbation.

Here in the right hand side we substituted the sensitivity of the
GOLF-instrument to the minimum fluid speed, $v_z=2~\rm{mm/sec}$, while
in the left hand side we substituted the frequency estimate
$\omega\sim N$ and the wave length through the CZ: $
R_{\odot}-z_{RZ}=3H=0.3R_{\odot}$.  For instance, substituting for the
magnetic field perturbation, $b_z/B_0=0.01$, $N=2.8\times 10^{-3}~
\rm{rad}~\rm{sec}^{-1}$ for the \Brunt frequency in the RZ,
$H=0.1R_{\odot}=7\times 10^9~\rm{cm}$~\citep{Bahcall88} for the
density scale height, one obtains $e^{-3K}/K\simeq 10^{-6}$, from
which the estimate $K=K_{max}\sim 4$ comes.

We organize our presentation as follows.  In
Section~\ref{section:basics} we formulate the MHD model for an ideal
plasma.  In Section~\ref{section:linear} we linearize the full set of
MHD equations and then derive a single master equation for the
z-component of the perturbation velocity $v_z^{(1)}(z)$. This
component leads to the Doppler shift of the optic frequencies measured
in helioseismic experiments.
In Subsection~\ref{subsection:B0zero} we check the validity of the
master equation against the well-known case of standard
helioseismology in an isotropic plasma, without magnetic fields. In
Subsection~\ref{subsection:B0} we derive the simplified master
equation in the perturbative limit. In this limit there are no MHD
(slow or Alfv\'{e}n) resonances within the Sun, a situation which was
treated in~\citep{Burgess04a}.
This perturbative method allows us to use a standard quantum
mechanical 1-D approach to determine an exact analytical spectrum of
g-modes in the presence of RZ magnetic fields.
In Section~\ref{section:discussion} we summarize our results.
 
\section{Basic ideal MHD equations}
\label{section:basics}
We describe the Sun as an ideal hydrodynamical system characterized by
nonlinear MHD equations. The mass conservation law for the total
density $\bar{\rho}$ can be written as
\begin{equation}\label{mass}
\frac{\partial \bar{\rho}}{\partial t} + \nabla\cdot(\bar{\rho}{\bf v})=0.
\end{equation}
The total pressure $\bar{\rho}$ is a sum of equilibrium, $\rho_0$, and 
non-equilibrium, $\rho$, parts, $\bar{\rho} =\rho_0 + \rho$.
Since viscosity can be neglected, momentum is conserved according to
\begin{equation}\label{momentum}
\frac{\partial \bf v}{\partial t} + ({\bf v}\cdot\nabla){\bf v}=
-\frac{1}{\bar{\rho}}\nabla \bar{P} + \frac{1}{4\pi \bar{\rho}}({\bf
B}\cdot\nabla){\bf B} + {\bf g}.
\end{equation}
Here the total pressure $\bar{P}=P_0(z) + P$ consists of two terms:
the equilibrium and non-equilibrium parts. The first is expressed as
$P_0(z)=p_0(z) + B_0^2/8\pi$ and obeys the equation $\nabla
P_0=\rho_0(z){\bf g}$.
The non-equilibrium part of the total pressure $P=p+ (1/4\pi)({\bf
B}_0\cdot {\bf B'})+ {\bf B'}^2/8\pi$ involves non-linear terms
coming from the total magnetic field ${\bf B}={\bf B}_0 + {\bf B'}$.

The evolution of the magnetic field is governed by Faraday's equation,
\begin{equation}
\label{Faradey}
\frac{\partial {\bf B}}{\partial t}= ({\bf B}\cdot \nabla){\bf
v}-({\bf v}\cdot \nabla){\bf B} - {\bf B}u,
\end{equation}
where $u=\nabla\cdot {\bf v}\neq 0$ is the compressibility of the gas.
Finally the conservation of entropy leads to the energy conservation
law for an ideal plasma,
\begin{equation}\label{energy}
\frac{\partial}{\partial
t}\left(\frac{\bar{p}}{\bar{\rho}^{\gamma}}\right)+ {\bf v}\cdot\nabla
\left(\frac{\bar{p}}{\bar{\rho}^{\gamma}}\right)=0,
\end{equation}
where $\bar{p}=p_0 + p$ is the total gas pressure.

\section{Linear MHD master equation}
\label{section:linear}
For definiteness here we consider the same generalized helioseismic
model already proposed in Ref.~\citep{Burgess04a}, adopting an
approximately rectangular rather than cylindrical, geometry.  In this
case it is convenient to choose a Cartesian coordinate system whose
z-axis is the ``radial''direction; opposite to the local acceleration,
${\bf g}=(0,0, -g(z))$. With this choice we take $z$ in the range $(0,
R_{\odot})$, where $z=0$ represents the solar center and $z=R_{\odot}$
denotes the solar surface. The radiative zone corresponds to $z\lsim
0.7R_{\odot}$.

The model assumes the background magnetic field to be directed along
the x-axis ${\bf B_0}=(B_0(z), 0,0)$, ensuring that physical gradients
lie along the z-axis. 
Such a field mimics both a toroidal field lying in the equatorial
plane and a poloidal field perpendicular to the equatorial plane.

We now linearize the above MHD equations (\ref{mass}-\ref{energy}), so
that all variables are split into background and fluctuating Euler
quantities, $f=f_0 + f^{(1)}$, with $f^{(1)}$ denoting small
fluctuations about the background value $f_0$. In what follows we also
neglect rotation of the Sun, so that the background velocity is zero,
$v_0=0$.
Moreover, we assume that the MHD perturbations
are independent of $y$, which implies that $B_y^{(1)}= v_y^{(1)}=0$,
leaving just six perturbation functions, instead of eight.

In addition, following \citep{Ruderman97}, we consider plane waves
propagating along the $x$-axis and therefore the dependence of all
functions on the coordinates $(x,z,t)$ can be reduced to just $\theta
\equiv x-Vt$, with $V=\omega/k_x$ being the phase velocity. This can
be done since all functions depend only on the harmonic factor
$e^{ik_xx - i\omega t}$.

This way all perturbations are functions of two variables,
$f^{(1)}(\theta, z)$, and the linearized MHD equations involve partial
derivatives over $\theta$ and $z$. Notice that $\partial/\partial t=
-V\partial/\partial \theta$ and $\partial /\partial x=
\partial/\partial \theta$.

Let us rewrite the initial system for the six functions
$f^{(1)}(\theta,z)$ in Eqs. (\ref{mass}-\ref{energy}) expressing all
of them through just two quantities, $v_z^{(1)}(z,\theta)$ and
$P^{(1)}(z,\theta)$. One obtains:
\begin{equation}\label{Eulerz1}
\frac{\partial P^{(1)}}{\partial z}=\frac{\rho_0D_A}{V}\frac{\partial
v_z^{(1)}}{\partial \theta} - \rho^{(1)}g,
\end{equation}
\begin{equation}\label{Eulerx1}
\frac{\partial v_z^{(1)}}{\partial z}=\frac{V}{F}\frac{\partial
P^{(1)}}{\partial \theta} + \frac{gV^2}{D_C}v_z^{(1)},
\end{equation}
\begin{equation}\label{energy2}
\frac{\partial \rho^{(1)}}{\partial
\theta}=\frac{V^2}{D_C}\frac{\partial P^{(1)}}{\partial \theta}
+\frac{v_z^{(1)}}{V}\frac{\rm{d}\rho_0}{\rm{d}z} +
\frac{\rho_0gD_A}{D_CV}v_z^{(1)}.
\end{equation}
\begin{equation}\label{B_x}
\frac{\partial B_x^{(1)}}{\partial \theta}=\frac{B_0(V^2
-c_s^2)}{\rho_0D_C}\frac{\partial P^{(1)}}{\partial
\theta}+\frac{v_z^{(1)}}{V}\frac{\rm{d}B_0}{\rm{d}z} +
\frac{gB_0V}{D_C}v_z^{(1)},
\end{equation}
\begin{equation}\label{B_z}
B_z^{(1)}=-B_0\frac{v_z^{(1)}}{V},
\end{equation}
\begin{equation}\label{v_x}
\frac{\partial v_x^{(1)}}{\partial
\theta}=\frac{Vc_s^2}{\rho_0D_C}\frac{\partial P^{(1)}}{\partial
\theta} - \frac{gv_A^2}{D_C}v_z^{(1)}.
\end{equation}
We introduced above the important coefficients $F,~D_A,~D_C$ which in
turn are functions of ``z'' via $\rho_0(z)$, the Alfv\'en velocity $v_A(z)$,
and the sound velocity $c_s(z)$:
\begin{eqnarray}\label{coefficients}
&&
F=\frac{\rho_0D_C}{V^2 - c_s^2}, \quad
D_A=V^2 -v_A^2, \nonumber\\ &&
D_C=(v_A^2+c_s^2)(V^2-c_T^2),
\end{eqnarray}
where $c_T^2=v_A^2c_s^2/(v_A^2 + c_s^2)$ is the squared cusp velocity.
The zeros of the last coefficient $D_C$ correspond to the slow
($V=c_T$) or Alfv\'{e}n ($V\approx v_A$, $v_A\ll c_s$) resonances
respectively.

Differentiating Eq.~(\ref{Eulerz1}) over $\theta $ and using the
next two equations one gets the master equation for the function
$v^{(1)}_z(\theta ,z)$ as:
\begin{eqnarray}\label{master}
&&\frac{\partial }{\partial z}F\frac{\partial }{\partial z}v_z^{(1)}-\rho
_{0}D_{A}\frac{\partial^{2}v_z^{(1)}}{\partial \theta ^{2}}- 
\nonumber\\
&&-\left[g^2\frac{F}{D_C} + \rho_0\frac{{\rm d}g}{{\rm d}z} + 
g\frac{{\rm d}}{{\rm d}z}\left(\frac{Fc_s^2}{D_C}\right)
\right]v_z^{(1)}=0. 
\end{eqnarray}
It is easy to see that, neglecting gravity, our generalized
equations~(\ref{Eulerz1})-(\ref{coefficients}) recover the total
system (34)-(38) derived in~\citep{Ruderman97} and Eq.~(\ref{master})
coincides with Eq.~(39) in~\citep{Ruderman97}.
In case of the constant gravity, $g$=const, while the sound velocity
$c_s^2(z)=\gamma gH(z)$ depends on the varying height scale $H(z)$,
Eq.~(\ref{master}) recovers Eq.~(7) in~\citep{Roberts}.

For simplicity we consider below the particular case of uniform and
constant magnetic field, $B_0$=const, constant gravity, $g$=const, and
density scale height, $H$=const.  Therefore the sound speed $c_s$ and
the \Brunt frequency $N$ are also constants in RZ, $c_s^2=\gamma
gH$=const, $N^2=(g/H)(\gamma -1)/\gamma$=const.  This is somewhat of a
simplification of the real Sun, but it allows us to derive qualitative
estimates of the magnetic field corrections to the pure g-mode
spectrum.
 
In what follows in our numerical estimates we will use the following
values: $\gamma=5/3$ (hydrogen plasma), $N=2.8\times 10^{-3}~
\rm{rad}~\rm{sec}^{-1}$, $H=0.1R_{\odot}=7\times
10^9~\rm{cm}$~\citep{Bahcall88}, hence $g=1.37\times
10^5~\rm{cm}~\rm{s}^{-2}$, $c_s=4\times 10^7~\rm{cm/s}$.

Separating in master equation~(\ref{master}) the exponential
dependence $e^{ik_{x}\theta }$ and taking into account the background
density profile $\rho _{0}=\rho _{c}e^{-z/H}$, one gets the ordinary
second order equation\footnote{Here we have corrected the factor in
  front of the first derivative in Eq.~(16) of~\citep{Burgess04a}
  changing $(\gamma-1)/\gamma H \to 1/H$. This does not affect the MHD
  spectra found there.}:
\begin{eqnarray}
\label{masterz1}
&&\left[1 - \frac{v_A^2}{V^2}\left(1 -
\frac{V^2}{c_s^2}\right)\right]\frac{{\rm d}^2v_z^{(1)}}{{\rm d}z^2}
-\frac{1}{H} \frac{{\rm d}v_z^{(1)}}{{\rm d}z} +\nonumber\\
&&+k_x^2\left(\frac{N^2}{\omega^2} - \left(1-\frac{V^2}{c_s^2}\right)\left[1 -
\frac{v_A^2}{V^2}\right]\right)v_z^{(1)}=0.
\end{eqnarray}
This equation generalizes Eq.~(16) of our work~\citep{Burgess04a} 
accounting for the compressibility parameter
$a_1=V^2/c_s^2=\omega^2/k_x^2c_s^2$($=0.24\omega^2/K^2N^2$ for hydrogen plasma
with $\gamma=5/3$). 
Note that $a_1$ is small for very low frequencies of g-modes $\omega\ll N$, 
$a_1\ll 1$, and therefore it was neglected in the problem considered 
in~\citep{Burgess04a}.

\subsection{Zero magnetic field limit}
\label{subsection:B0zero}
For $v_A=B_0=0$ with the use of the transformation
$v_z^{(1)}(z)=\Psi(z)e^{z/2H}$ Eq.~(\ref{masterz1}) converts into
\begin{equation}
\label{masterz0}
\frac{{\rm d}^2\Psi}{{\rm d}z^2} + \left[\frac{\omega^2}{c_s^2} - \frac{1}{4H^2} 
+ k_x^2\left(\frac{N^2}{\omega^2} -1\right)\right]\Psi=0.
\end{equation}
This 1-D MHD equation coincides with the 
3-D oscillation Eq.~(7.90) in~\citep{Christensen03}:
\begin{equation}
\label{masterz3D}
\frac{{\rm d}^2\Psi}{{\rm d}r^2} + \left[\frac{\omega^2}{c_s^2} -
\frac{1}{4H^2} + k^2_h\left(\frac{N^2(r)}{\omega^2}
-1\right)\right]\Psi=0,
\end{equation}
here we have substituted the acoustical cut-off frequency
$\omega_c=(c_s^2/4H^2)(1-2dH/dr)\approx c_s^2/4H^2$ and put
$S_l^2=l(l+1)c_s^2/r^2\simeq k^2_hc_s^2$, see Eq.~(4.60)
in~\citep{Christensen03}.

We would like to stress that, in the JWKB approximation for low 
frequency g-modes, $c_s^2\gg V^2$, $S_l^2\gg \omega^2$, 
both equations (\ref{masterz0}) and 
(\ref{masterz3D}) lead to the analogous spectra:
\begin{eqnarray}\label{JWKB}
&&\omega_g(n,k_x)=\frac{k_xN}{\pi
(n+1/2)}z_{RZ}~~~\mbox{(1D~model)},\nonumber\\
&&\omega_g(n,l)=\frac{\sqrt{l(l+1)}\int_{r_1}^{r_2}N(r)dr/r}{\pi (n +
l/2 + \alpha_g)}~~~\mbox{(3D model)},
\end{eqnarray} 
where $n$ is the radial order of the wave, and the transversal wave number
$k_x$ is the 1-D analogue of the angular degree $l$ in the 3-D case.

Let us denote the factor in brackets in Eq.~(\ref{masterz0}) as
$\beta^2/4H^2$,
\begin{equation}\label{beta0}
\beta=\sqrt{4K^2\left(\frac{N^2}{\omega^2} -1\right) - 1 +4K^2a_1},
\end{equation}
here $K=k_xH$ is the dimensionless wave number, $K\geq K_{min}\simeq 2\pi
H/R_{\odot}\sim 1$.
Eq.~(\ref{masterz0}) can be rewritten using Eq.~(\ref{beta0}) as
\begin{equation}\label{masterz00}
\frac{{\rm d}^2\Psi}{{\rm d}z^2} + \frac{\beta^2}{4H^2}\Psi (z)=0.
\end{equation}
In RZ where $N_{RZ}\equiv N=const\neq 0$ 
one obtains
the solution of Eq. (\ref{masterz00}) in the form $\Psi_{RZ}
(z)=C_{RZ}\sin (\beta z/2H)$, where $C_{RZ}$ is a constant. 
It accounts for the 
boundary condition $v_z^{(1)}(0)=0$ at the center of the Sun. 
In CZ we approximate
\Brunt frequency by the value $N_{CZ}=0$.  In that case $\beta$ tends to
$i\Gamma=i\sqrt{4K^2(1-a_1) + 1}$. Taking into account  
a second boundary condition, that there are no solutions which grows with $z$,  
one gets the decaying MHD wave in the form: 
$\Psi_{CZ}(z)=C_{CZ}\exp (- z\Gamma/2H)$. 

Now matching both solutions at
the top of RZ, $z=z_{RZ}$, (for logarithmic derivatives see the
book by~\citet{LL}), one obtains the dispersion equation in the case
$B_0=0$:
\begin{equation}\label{dispersionLL}
\beta \cot \left[\frac{\beta z_{RZ}}{2H}\right]= -\Gamma,
\end{equation}
or the g-mode spectrum in our 1-D model is given by 
$\beta z_{RZ}/2H+\pi n= \arccot [-\Gamma/\beta]$, $n= -1, -2, -3,...$
(see solid curves in the Fig. 1). 
\begin{figure}
\includegraphics[width=0.95\columnwidth]{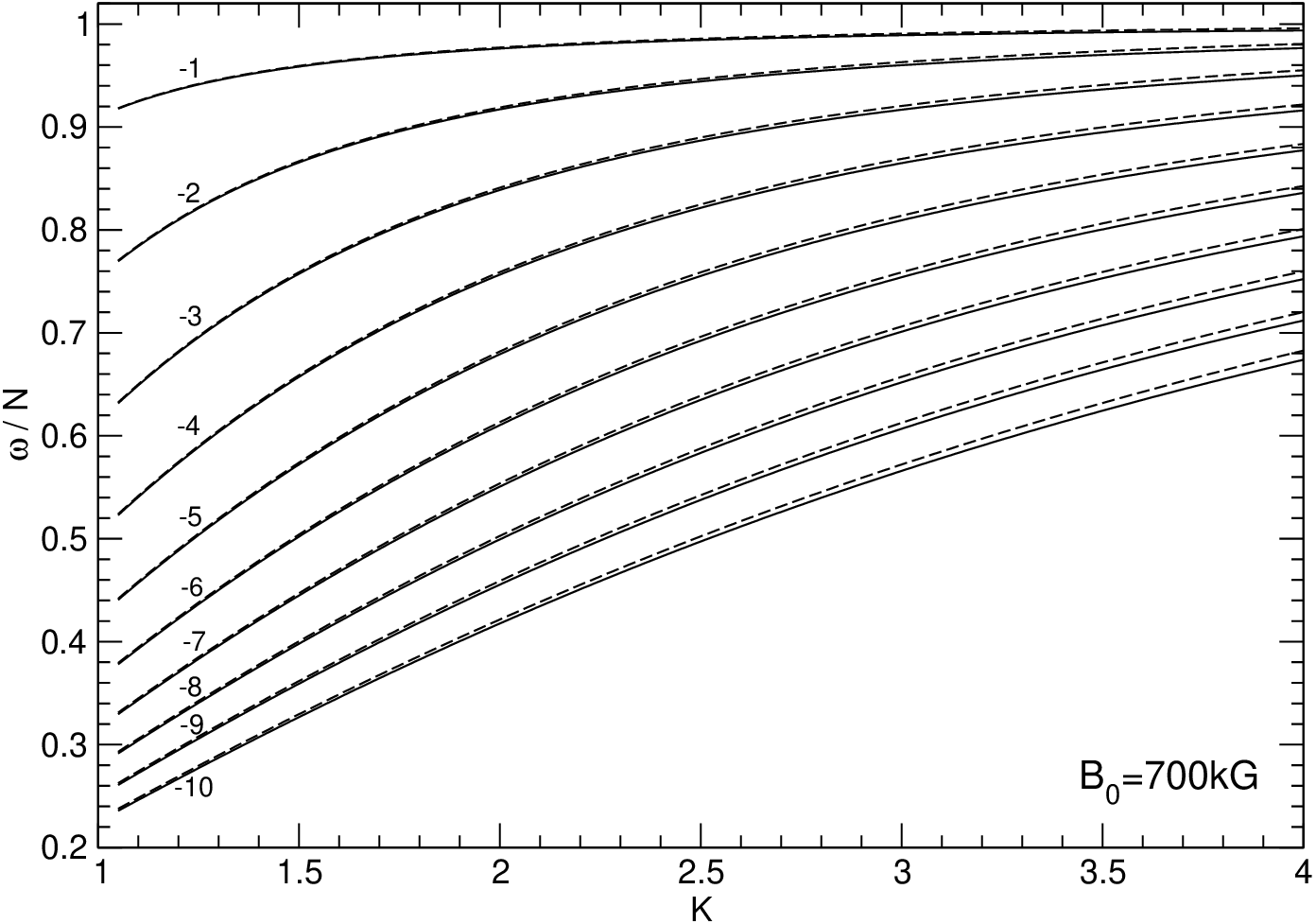}
\caption{\label{fig:freq-K} The g-mode frequencies for
  $n=-1,-2,\dots,-10$ normalized to \Brunt frequency, $\omega/N$,
  versus wave number, $K$. Solid lines are for zero magnetic field,
  while dashed lines correspond to a fixed magnetic field
  $B_0=700$kG.}
\end{figure}

\subsection{Magnetic corrections to the g-mode spectrum}
\label{subsection:B0}

In order to obtain spectra of g-modes in the presence of magnetic
field let us define the coefficient in front of the second order
derivative in Eq.~(\ref{masterz1}) as $1-\zeta$, where
$\zeta=v_{A0}^2e^{z/H}(1-a_1)/V^2$. The Alfv\'{e}n velocity at the
solar center is given by
$v_{A0}=B_0/\sqrt{4\pi \rho_c}\approx(B_0/43.4~\mbox{Gauss})~\mbox{cm~sec}^{-1}$,
here, and through the paper, $\rho_c=150~\mbox{g/cm}^3$.

\begin{figure}
\includegraphics[width=0.95\columnwidth]{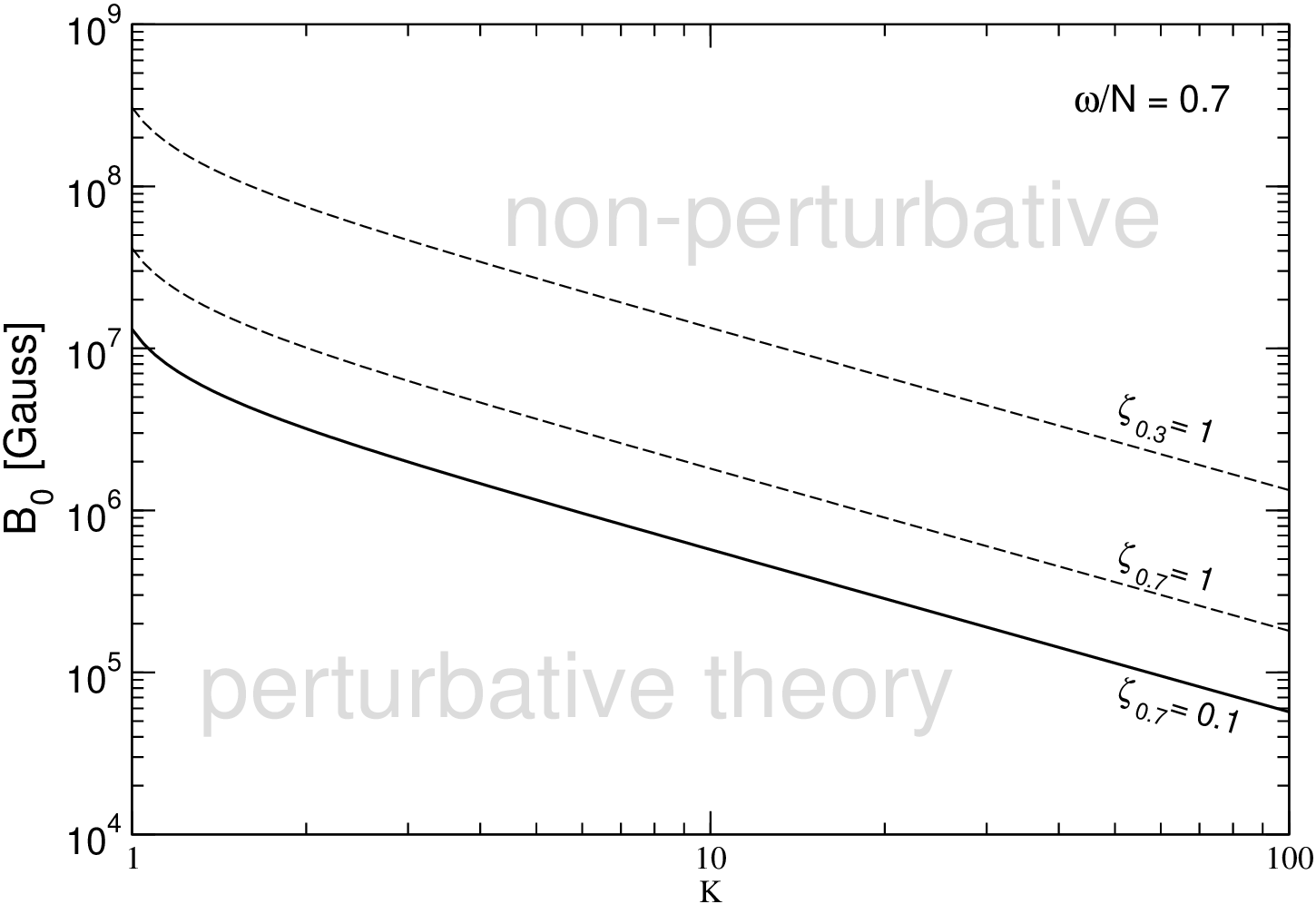}
\caption{\label{fig:B0-K} The resonance positions are shown in the
  plane $B_0-K$ (magnetic field {\it vs} wave number along magnetic
  field). The curves separate the regions where perturbative or
  non-perturbative approaches may or should be used. $\zeta_z=1$ means
  that at place $z$ the Alfv\'en resonance occurs.}
\end{figure}

We consider the perturbative regime for magnetic fields, where
$v_A^2=v_{A0}^2e^{z/H}\ll V^2=\omega^2/k_x^2=
[N^2H^2/K^2](\omega/N)^2$, so that $\zeta\ll 1$ and MHD resonances
($\zeta=1$) do not appear within the RZ. This region is the lower one
in Fig.~\ref{fig:B0-K}. The dashed curves labelled $\zeta_{0.3}= 1$
and $\zeta_{0.7}= 1$ correspond to resonances that occur at
$0.3R_{\odot}$ and $0.7R_{\odot}$ in the non-perturbative region. The
solid curve is chosen to illustrate the separation between the two
regimes, according to the criterion
$\zeta_{0.7}=\zeta(0.7R_\odot)=0.1\ll 1$.
The perturbative magnetic field for which the maximum Alfv\'{e}n
velocity is small, $v_A^2(0.7R_{\odot})\ll V^2$, obeys
\begin{equation}\label{perturb}
B_0\ll \frac{2.8\times 10^7}{K}~\mbox{Gauss} \: (\rm for  \: \:  \omega\lsim N).
\end{equation}

The absence of MHD resonances allows us to set
\[
v_z^{(1)} \equiv \left[1 - \frac{v_A^2}{V^2}\left(1-\frac{V^2}{c_s^2}\right)\right]^{-1/2}e^{z/2H}\Psi (z),
\] 
and thus derive from Eq.~(\ref{masterz1}) the
$O(v_A^2/V^2)$-correction to Eq.~(\ref{masterz00}):
\begin{equation}\label{perturbation}
\frac{\rm{d}^2\Psi}{\rm{d}z^2} + \left[\frac{\beta^2}{4H^2}+ k_B^2\right]\Psi (z)=0,
\end{equation}
here the wave number correction $k^2_B$ is given by
$$
k_B^2=(1-a_1)\frac{v_A^2}{V^2}k_x^2\left[\frac{N^2}{\omega^2} +a_1\right].
$$
Introducing the notation:
\begin{equation}\label{A0}
A_0=(1-a_1)\frac{v_{A0}^2}{V^2}\left[\frac{N^2}{\omega^2} +a_1\right]>0,
\end{equation}
one gets from Eq.~(\ref{perturbation}):
$$ \frac{\rm{d}^2\Psi}{\rm{d}z^2} + \left[\frac{\beta^2}{4H^2}+
k_x^2A_0e^{z/H} \right]\Psi (z)=0
$$
and after the change $2s=z/H + \ln (4K^2A_0)$ this equation takes the form
\begin{equation}\label{Bessel}
\frac{{\rm d}^2\Psi}{{\rm d}s^2} + [\beta^2 + e^{2s}]\Psi (s)=0.
\end{equation}
The general solution of Eq.~(\ref{Bessel}) is expressed via the Bessel
functions of the first kind, $\Psi (s)=C_1J_{i\beta}(e^s) +
C_2J_{-i\beta}(e^s)$ (see~\citet{GR}), $C_1$ and $C_2$ are constants.
Now, coming back to the variable $z$, we use the boundary condition at
the solar center, $\Psi (0)=0$ ($v_z(0)=0$), to obtain the solution of
Eq.~(\ref{Bessel}) in RZ, $0\leq z\leq z_{RZ}$,
\begin{eqnarray}\label{perturbation1}
&&\Psi_{RZ} (z)=C_{RZ} \left[J_{i\beta}(2KA_0^{1/2}e^{z/2H}) -\right.\nonumber\\
&&\left.-\frac{J_{i\beta}(2KA_0^{1/2})}{J_{-i\beta}(2KA_0^{1/2})}
J_{-i\beta}(2KA_0^{1/2}e^{z/2H})\right].
\end{eqnarray}

In CZ the \Brunt frequency $N$ vanishes, $N=0$, so that we are led to
the same solution as in isotropic case $\Psi_{CZ}(z)=C_{CZ}\exp
(-z\Gamma /H)$, neglecting the possible existence of a CZ magnetic
field (if such a field is present, with strength of 300~kG, the g-mode
frequencies change by no more than $10^{-5}$).
Since the arguments of Bessel functions in Eq.~(\ref{Bessel}) are small, 
we can use the first terms in the Bessel function series only,
\begin{eqnarray}
&&J_{\nu}(z)=\left(\frac{z}{2}\right)^{\nu}\sum_{k=0}^{\infty}\frac{(-1)^k}{k!\Gamma
(\nu + k + 1)}\left(\frac{z}{2}\right)^{2k}, ~~~\mid
arg~z\mid<\pi,\nonumber\\ 
&&J_{\pm i\beta}(2KA_0^{1/2}e^{z/2H})\approx
\frac{(KA_0^{1/2}e^{z/2H})^{\pm i\beta}}{\Gamma (1 \pm i\beta)}\times
\nonumber\\&& \times\left[1 - \frac{K^2A_0e^{z/H}}{1\pm i\beta} +
O(K^2A_0e^{z/H})\right],
\end{eqnarray}
Then by matching the logarithmic derivatives of the solutions 
$\Psi_{RZ}$ and $\Psi_{CZ}$ at the top of RZ, $z=z_{RZ}$,
one obtains the generalized dispersion equation for the case $B_0\neq 0$:
\begin{eqnarray}\label{dispersionB0}
&&\beta\left(1 + \frac{2\kappa_{RZ}^2}{1 + \beta^2}\right)\left(1 -
\beta\frac{2\kappa_{RZ}^2}{(1 + \beta^2)} \frac1{\sin (\beta
z_{RZ}/H)}\right) \times\nonumber\\
&&\times\cot \left[\beta\frac{z_{RZ}}{2H}\right]-
\frac{2\kappa_{RZ}^2}{1 + \beta^2}= - \Gamma,
\end{eqnarray}
where $\kappa_{RZ}=KA_0^{1/2}e^{z_{RZ}/2H}$. 

This is our main equation. Clearly, when the magnetic field correction
tends to zero, hence $2\kappa_{RZ}^2/(1 + \beta^2)$ tends to $0$, one
recovers the isotropic case, given in Eq.~(\ref{dispersionLL}).

We search for a perturbative solution of Eq.~(\ref{dispersionB0}) of
the form $\beta=\beta_0(1 + \delta_B)$ and $\omega=\omega_0(n)(1 +
\alpha_B(n))$ where $\beta=\beta_0$ and correspondingly
$\omega=\omega_0(n)$ are the solution of Eq. (\ref{dispersionLL}) for
$B_0=0$ and the smallness of $\delta_B\ll 1$ and $\alpha_B(n)\ll 1$
follows from  $\kappa^2_{RZ}\ll 1$.
%
% In linear approximation we find
% %
% \begin{equation}\label{linearcorr}
% \delta_B= - \frac{2\kappa_{RZ}^2}{1 + \beta_0^2}\times\frac{[1/\Gamma
% + \Gamma/2\beta_0 + \beta_0/2\Gamma]}{(1+ (\beta_0z_{RZ}/2H)[
% \beta_0/\Gamma + \Gamma/\beta_0])}
% \end{equation}
% %
% and
% %
% \begin{equation}\label{linearshift}
% \alpha_B=-\delta_B\frac{\beta_0^2/4K^2}{N^2/\omega_0^2-0.24\omega_0^2/K^2N^2}.
% \end{equation}

\begin{figure}
\includegraphics[width=0.95\columnwidth]{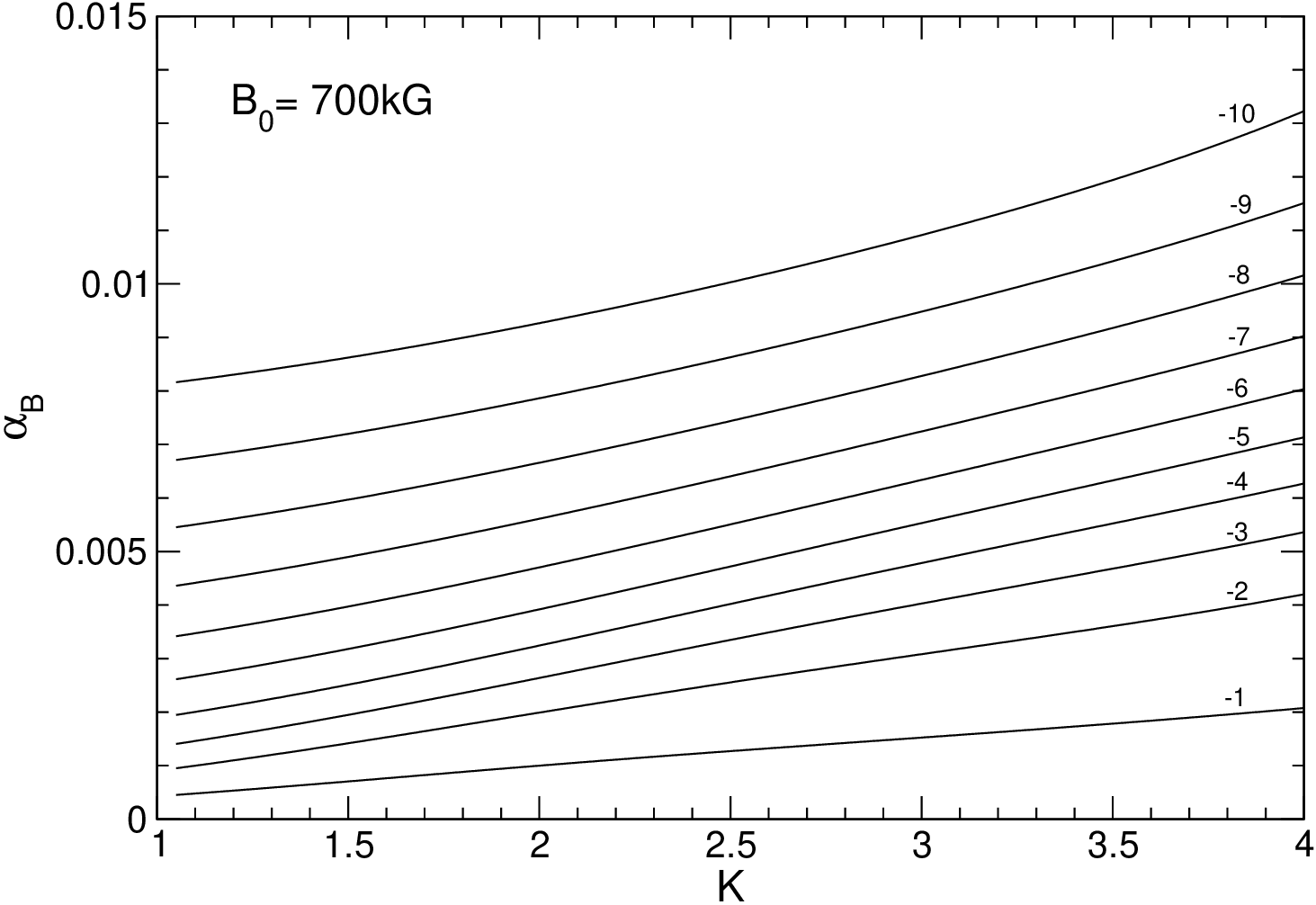}
\caption{\label{fig:alpha-K} The shift of g-mode frequency $\alpha_B$
  versus wave number, $K$, shown for fixed magnetic field strength,
  $B_0=700$~kG and different mode radial orders, $n=-1,-2,\dots,-10$.}
\end{figure}

\begin{figure}
\includegraphics[width=0.95\columnwidth]{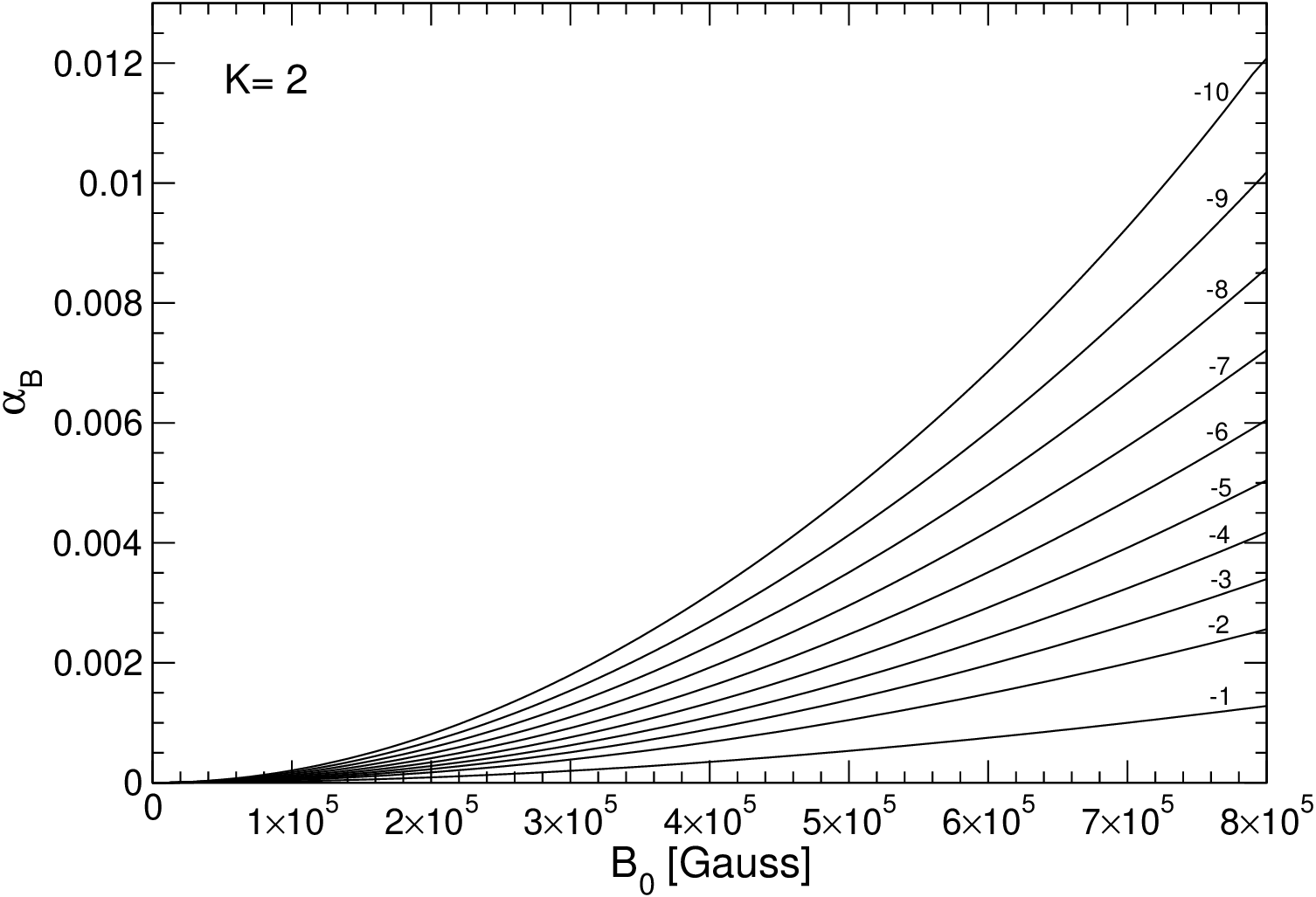}
\caption{\label{fig:alpha-B0} The shift of g-mode frequency $\alpha_B$
  versus magnetic field strength, $B_0$, shown for fixed wave number,
  $K=2$, and different mode radial orders, $n=-1,-2,\dots,-10$.}
\end{figure}

The g-mode spectra for fixed $B_0=700$~kG are shown on
Fig.~\ref{fig:freq-K} (see dashed curves).  Figs.~\ref{fig:alpha-K}
and ~\ref{fig:alpha-B0} display our results for $\alpha_B$ as a
function of mode radial order $n$, magnetic field and wave number $K$.
One sees that the magnetic field shift $\alpha_B$ of a g-mode
frequency $\omega_0(n)$ is always positive, $\alpha_B>0$.
In Fig.~\ref{fig:alpha-K} we show the absolute values of the shift
$\alpha_B(n)$ for different g-modes and for fixed $B_0=700$~kG as a
function of $K$. Conversely, in Fig.~\ref{fig:alpha-B0} we fix the
wave number $K=2$ and plot $\alpha_B(n)$ as a function of $B_0$.
One sees that, the higher the mode radial order $|n|$, the less the magnetic
field strength required to produce a given g-mode frequency shift.

\section{Discussion}
\label{section:discussion} 
We have given a generalization of helioseismology to account for the
presence of central magnetic fields in the Sun. We have determined the
resulting g-mode spectrum within the framework of a perturbative
one--dimensional Magneto-Hydrodynamics model.

There are three factors influencing g-mode observation in helioseismic
experiments. 
First, such g-modes should have long wave lengths to penetrate the CZ:
in our case $K\leq 4$, and in the 3D-model, low values of $l$.
Second, the radial order $n$ should also be low, otherwise, such low
frequency g-modes are much below the present experimental sensitivity.
Third, there is a strong influence of the RZ magnetic field. 

If the magnetic field is too strong (more than a few MG) all g-modes
are locked within the Alfv\'{e}n cavity (see~\citet{Burgess04a}),
hence these g-modes decay far beneath the CZ becoming invisible in the
photosphere.  This happens because the MHD energy in the radial
direction is fully diverted to the transversal plane at the Alfv\'{e}n
resonant layer position. On the other hand, for very small magnetic
fields ($B_0\ll 1$ MG) the magnetic field effect is negligible and can
not account for a possible discrepancy between present experimental
data and theoretical predictions of seismic models. In contrast, we
have shown that a solar radiative zone magnetic fields of intermediate
magnitude, of the order 600-800~kG, can displace the pure g-mode
frequencies by about 1\% with respect to our model of seismic wave
propagation, a value close to what is hinted by results of the GOLF
experiment.

We find that higher modes require a smaller magnetic field to produce
a given g-mode frequency shift.
However encouraging this result may sound, let us stress again that in
our simple one-dimensional MHD picture (with further
assumptions, such as adiabatic oscillations, exponential density
profile, constant gravity, etc) we can only make a qualitative
estimate of the magnetic field corrections to the pure g-mode
spectrum. For example, our formalism can not explain the magnetic
splitting of g-mode frequencies over azimuthal number $m$, as it
requires 3-D.
Further work in 3-D geometry is necessary to perform a quantitative
comparison with the frequency patterns observed in the GOLF experiment
\citep{TC04}.  Even within the simple analytic approach 1-D MHD model
one may include viscosity effects in non-ideal plasma with finite
conductivity, and also take into account magnetic field diffusion
stabilizing MHD instabilities.

Last, but not least, recall that our perturbative analysis avoids the
appearance of MHD resonances that could lead to density spikes. These
are potentially important, as they can affect neutrino propagation
through the solar RZ~\citep{Burgess03,Burgess04a,Burgess04b}.
Improved determination of neutrino mixing parameters, e.g. by
KamLAND~\citep{KamLAND}, allows one to carry out neutrino tomography
in deep solar interior.
Both regimes of ``magnetized helioseismology'' (i) the MHD seismic
models and (ii) the analysis of MSW neutrino oscillations in noisy Sun
are complementary tools to explore RZ magnetic fields. A fully
quantitative analysis may require the inclusion of the differential
rotation in the RZ~\citep{TC05} as well as non-linearities.

\vskip .2cm

Note added: As this paper was being typed, we saw a paper by
\citet{Hasan05} (astro-ph/0511472), where a similar idea to probe the
internal magnetic field of slowly pulsating B-stars through g modes is
given.  Their 3-D results are consistent with our simpler 1-D
estimates.

\section*{Acknowledgments}

Work supported by Spanish grant BFM2002-00345. We thank N.~Dzhalilov,
M.~Ruderman and S.~Turck-Chi\`eze for discussions.  TIR and VBS thank
IFIC's AHEP group for hospitality during part of this work. They were
partially supported by RFBR grant 04-02-16386 and the RAS Presidium
Programme ``Solar activity''. TIR was supported by a Marie Curie
Incoming International Fellowship of the EC.

 %%%%%%%%%%%%%%%%%%%%%%%%%%%%%%%%%%%%%%%%%%%%

\label{lastpage}

\end{document}